\documentclass[12pt]{spieman}

\usepackage{amsmath,amsfonts,amssymb,aas_macros,lineno}
\usepackage{graphicx}
\usepackage{setspace}
\usepackage{tocloft}
\usepackage[colorlinks=true, allcolors=blue]{hyperref}

\title{SMA-X: Versatile information sharing in and around telescopes, and beyond}

\cftpagenumbersoff{figure}
\cftpagenumbersoff{table}

\begin{document} 

\author[a,*]{Attila Kov\'{a}cs}
\author[a]{Paul K. Grimes}
\author[a]{Christopher Moriarty}
\author[a]{Robert Wilson}
\affil[a]{Center for Astrophysics $|$ Harvard \& Smithsonian, 60 Garden St, Cambridge, MA 02138, U.S.A.}

\maketitle

% Include email contact information for corresponding author
{\scriptsize \noindent \textbf{*}Address all correspondence to Attila Kov\'{a}cs, \linkable{attila.kovacs@cfa.harvard.edu} }

\begin{abstract}
We developed the SMA eXchange (SMA-X) as a real-time data sharing solution, built atop a central Redis database. SMA-X is a storage convention, facilitated by a set of server-side Lua scripts (or Redis functions) which enable efficient low-latency and high-throughput real-time sharing of hierarchically structured data among the various systems and subsystems of the telescope. It enables fast, atomic retrievals of specific leaf elements, branches, and sub-trees, including associated metadata (types, dimensions, timestamps, and origins, and more). At the Submillimer Array (SMA) we rely on it since 2021 to share a diverse set of $\sim$10,000 real-time variables, including arrays, across more than 100 computers, with information being published every 10\,ms in some cases. SMA-X is open-source, and is freely available to everyone through a set of public GitHub repositories, including C/C++ and Python3 libraries to allow integration with observatory applications. A set of command-line tools provide access to the database from the POSIX shell and/or from any scripting language, and we also provide a configurable system daemon for archiving the observatory state at regular intervals into a time-series Structured Query Language (SQL) database to create a detailed historical record.
\end{abstract}

{\scriptsize \noindent \copyright The Authors. Published by SPIE under a Creative Commons Attribution 4.0 International License. Distribution or reproduction of this work in whole or in part requires full attribution of the original
publication, including its DOI. [DOI: \href{https://doi.org/10.1117/1.JATIS.11.1.017001}{10.1117/1.JATIS.11.1.017001}] }


{\scriptsize \noindent Paper 24149G received Sep. 20, 2024; revised Dec. 26, 2024; accepted Dec. 30, 2024. }

%\linenumbers
%\begin{spacing}{2}   % use double spacing for rest of manuscript

\section{Introduction}
\label{sec:intro}  % \label{} allows reference to this section

When operating an observatory, the control system must gather information from hundreds of sources in quasi real time: telescope drive encoders, sensors from various equipment and devices, diagnostic data from instrumentation, weather data, various programs running on different computers etc. At the Submillimeter Array (SMA)\cite{SMA}, we monitor data from 8 antennas, 9 weather stations, 16 receivers, 48 correlator units, dozens of control computers, hundreds of programs and sensors. The distributed programs of the control system must have access to a selection of the shared information to manage observations, operate hardware optimally, and to create fully-described scientific data products. When components report values that are outside of their normal operating ranges, it is important to act without delay, both so that affected data can be flagged appropriately for problems, and to bring it to the attention of operators, who may be able to correct it. How well information is shared among systems in a telescope is a major factor in determining the efficiency of its operation. Even small latencies can compound in the chain of information flow and result in significant idle times, or else manifest in degraded telescope performance and/or poor data quality.

\begin{figure}[!hbt]
\centering
\includegraphics[width=0.80\textwidth]{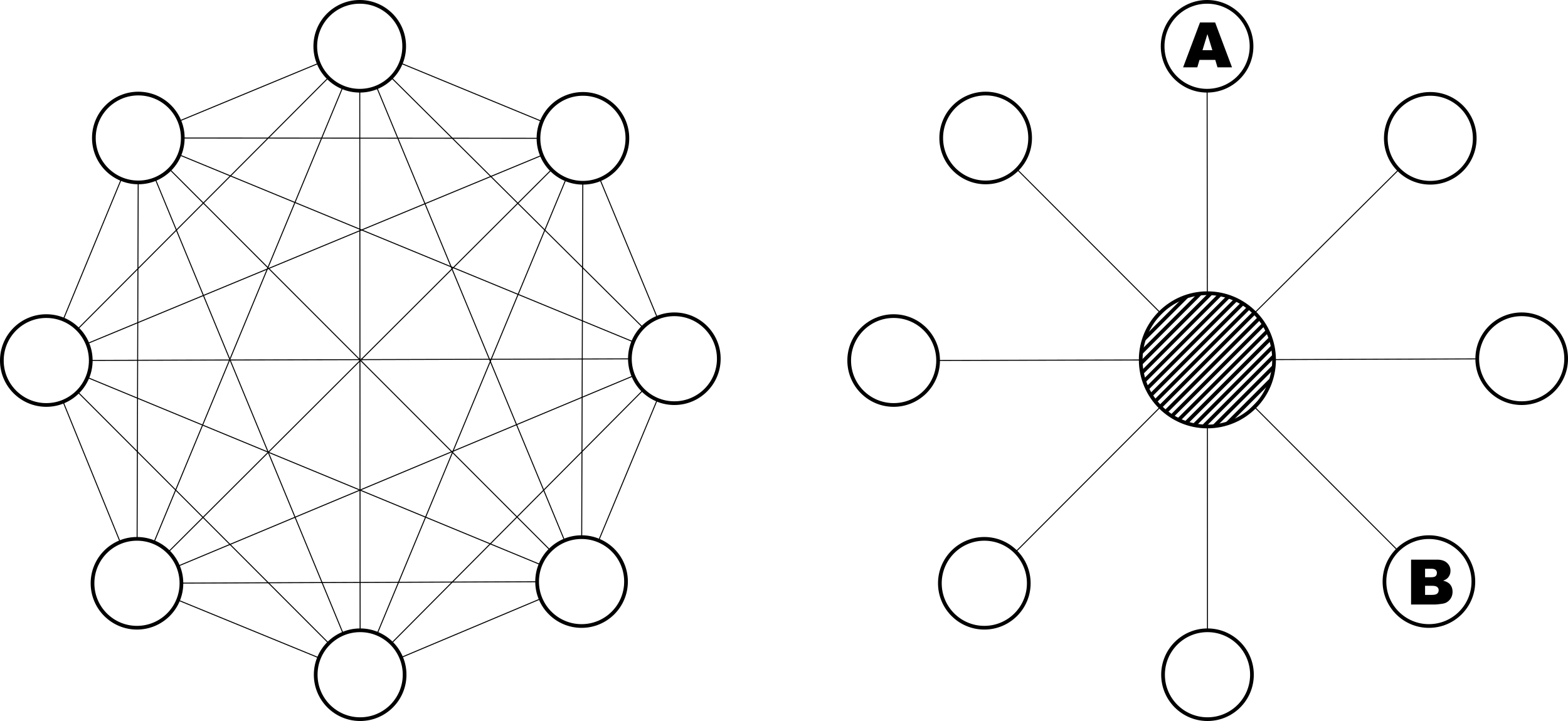}
\caption{\label{fig:diagram} {\em Left:} Maximal network of decentralized P2P sharing among nodes, s.t. each node has access from information from all other nodes. {\em Right:} Centralized information sharing. The same graph may represent P2P sharing to a central (shaded circle) "control system" or middleware node, or else to a public message board. In the P2P case, nodes A and B are isolated from one another and do not have access to each other's shared data, unless the central control system explicitly forwards information from one node to another -- decision making must be predominantly centralized. In the case of the central message board, however, nodes A and B have unfettered access each other's shared data, as well as data shared from any other node, so they can make their own control decisions as appropriate. A central message board, like SMA-X, enhances information sharing and enables decentralized decision making with a minimal network complexity.}
\end{figure}

Different telescopes have found different answers to information sharing for their monitoring and control needs\cite{Carrasco2024}. A number of observatories rely on peer-to-peer (P2P) communication, typically through ethernet (e.g.\ TCP/IP or UDP sockets), or other forms of digital links. By P2P, we define any communication between a pair of endpoint nodes, in which messages produced at point 'A' are explicitly addressed to the consumer point 'B', or else explicitly requested by point 'B' from the address of point 'A' --- regardless of whether the transaction is direct or via some middleware (see Figure~\ref{fig:diagram}).

Until recently, the SMA has relied almost exclusively on P2P information sharing both over commercial reflective memory hardware, and using Open Network Computing (ONC) Remote Procedure Calls (RPC)\cite{RPC}. Other telescopes, like APEX, developed their own custom P2P messaging systems\cite{APECS}, e.g.\ with communication over UDP sockets.

\section{Beyond P2P}

Such P2P systems have clearly been reasonably successful in providing the necessary means of communication among telescope systems. In some cases P2P is used primarily in relation to a centralized decision making 'control-system' software or middleware (such as APECS\cite{APECS,APECS-next} at the APEX telescope). Alternatively, P2P links may route specific bits of information between various dependent programs of a distributed control system (see Fig.~\ref{fig:diagram}). However, P2P information sharing has at least two fundamental flaws: inflexibility and unnecessary complexity.

What happens when a new instrument, component is added to the system? Or, if a program is changed to have new switches, or it produces different / new values than it did before; or is migrated from one IP address to another (heavens forbid to a different network even)? The P2P system of information sharing does not usually handle such changes well. For every change, several programs may have to be explicitly modified to send or receive vital information regarding the new components, or to route data to/from a new address. Centralized control systems, such as APECS, may have to be re-initialized every time the configuration changes relating to how or what information is shared in the telescope ecosystem. Because control systems typically handle information from multiple sources in parallel, every such change has the potential to create new race conditions, bottlenecks, or even deadlocks. Such unexpected side-effects from changes are often non-trivial to trace and/or fix, especially because the developer too can only peek at the shared information through the programs that are on either end of individual P2P links.

A second, equally nagging issue of P2P systems, is that the number of P2P links may scale as $\mathcal{O}(N^2)$ for $N$ information sharing nodes present in a distributed control systems. Thus, when telescopes have hundreds of individual components that produce and/or consume shared information, the number of P2P links can easily grow into the thousands. Consequently, information flow can become tedious to trace and manage. When something does not work as expected, it may take a very long time to track down how related bits and pieces of information criss-crossed their way across the system, and where it all went wrong exactly, especially because the information exchange happens on private channels, which do not facilitate external diagnostics.

Thus, while P2P systems can be pseudo-stable in an unchanging telescope environment, they all too often become visibly unstable in a constantly evolving observatory. As most observatories are in a perpetual state of development, perhaps P2P is not an ideal solution for them, in general.

\subsection{Public Information \& Decentralized Decision Making}

One may observe that every bit of data is either produced or consumed by a particular component in the system, regardless of how complex that system is. There is no reason why the {\em producers} of information should prejudice who or what programs can access their shared data (or how). A weather station that measures humidity, does not need to have a definitive list of all programs in the telescope that need a humidity value. Conversely, a program that needs a humidity value, should also not care what specific component or program, at what IP address, provides the value. But that is exactly what P2P requires at its core, P2P dictates that either the producer of information send it to its consumer(s) explicitly, or else that the consumer(s) of information request it from the specific producer. This is the case even if middleware is involved in the (re)distribution of information -- which may save the weather station from needing to know all consumers of weather data, only for the middleware to be needing that knowledge instead.

A better way is to share information without the explicit routing that P2P involves, such as via a public message board. Producers of information simply post their data to the public message board, and {\em any} consumer that needs {\em any} bit of shared information can obtain it from there. The general availability of information about all components is conducive to decentralized decision making also, as different programs may individually gather the information they need to make appropriate decisions on operating a particular sub-system without having to control all equipment from a central hypervisor middleware) software. A public message board can be implemented with a publish-subscribe (PUB/SUB) system (e.g.\ the one onboard Stratospheric Observatory for Infrared Astronomy [SOFIA]\cite{SOFIA}), and/or via a real-time database (such as Redis or memcached).

\subsection{PUB/SUB vs Persistent Real-time State}
Publish-subscribe refers to messaging systems, where information that is {\em published} to a specific {\em channel} is immediately forwarded to the set of active {\em subscribers} of that channel. For example, SOFIA's Mission Controls and Communication System (MCCS)\cite{SOFIA-HK} published housekeeping data at regular intervals (e.g.\ 1\,Hz, 10\,Hz, 50\,Hz) in this way. Depending on what cadence was required by the consumer program, it would subscribe to the appropriate {\em channel(s)}, and would receive updated telescope data 'streamed' to it at the desired rate. 

This works well for data that is broadcast frequently enough, such that it does not matter much if one misses an occasional update. If a consumer to the 1\,Hz 'stream' subscribed to a channel just after a message was sent on it, it will have to wait at most a second before it receives the next update. Some publish-subscribe systems (but not all) overcome that limitation by sending the last message to new clients upon connecting. However, while this ensures that clients are state aware immediately after connection, the client might not know how long ago that last message originated, and hence whether this initial state is valid or else stale.

The second public message board solution is to use a real-time database instead, which effectively stores a snapshot state of all shared variables at any given time. Producers of information update their state variables when these change, and consumers can poll them when they need it. However, this too can be abused easily. One may flood the network with high-frequency polling of slowly changing data, or else suffer unnecessary delays if polling data less frequently.

This is why we developed the SMA eXchange (SMA-X). It offers the best of both worlds: it can be used to provide the most current state of data, together with origination timestamps and other metadata, regardless of when a client connects, and it also notifies clients immediately when variables of interest are updated, without the need for frequent polling to detect changes in state -- all with minimal network traffic to conduct transactions.

\section{The SMA eXchange (SMA-X)}

The SMA eXchange (SMA-X) is a data storage and messaging convention built on top a central Redis database (version 4 or later). SMA-X is facilitated by a set of server side Lua scripts (or functions), which provide a higher-level interface to the Redis or Valkey database. Redis is primarily a very efficient open-source key/value storage solution, which includes a PUB/SUB subsystem also. Data are both labeled and stored verbetum as 'strings'. Beyond that, Redis and its clones like Valkey or Dragonfly, offer features that make them an ideal choice for a data sharing solution for distributed systems, in general:

\begin{itemize}
\item{Stored data can be grouped into tables (hash tables in Redis terminology), which may be accessed efficiently and atomically both in their entirety or by individual fields.}
\item{Supports pipelined (batch) mode access.}
\item{Includes a PUB/SUB subsystem.}
\item{Allows server-side scripting (via the Lua language) to implement higher-level atomic data access.}
\end{itemize}

\begin{table}[htb]
\caption{SMA-X basic storage types}
\centering
\begin{tabular}{|c|l|}
\hline
 {\bf Data Type} & {\bf Description} \\
\hline
{\ttfamily int[8,16,32,64]}     & signed integer value(s), with or without specific widths, e.g.\ {\ttfamily int} or {\ttfamily int32} \\
\hline
{\ttfamily float}               & single-precision (32-bit) floating point value(s) \\
\hline
{\ttfamily double}              & double-precision (64-bit) floating point value(s) \\
\hline
{\ttfamily boolean}    	        & logical true/false value(s) \\
\hline
{\ttfamily string}   	        & ASCII string value(s), stored in JSON-style\cite{JSON, JSON-ISO} escaped form \\
\hline
{\ttfamily struct}    	        & reference to a Redis hash table containing sub-structure data \\
\hline
\end{tabular}
\label{tab:data-types}
\end{table}

\subsection{Feature Overview}

\subsubsection{Basic data types} 
Individual SMA-X variables are stored as 'fields' in Redis hash tables, either as scalar values or as flattened arrays. SMA-X supports all primitive and numerical data types used in typical programming languages. Individual variables are stored in serialized (ASCII) representation which allows platform independent access to these. See Table~\ref{tab:data-types} for details.

\subsubsection{Metadata} 
For every variable stored in SMA-X, we also store a set of essential metadata, such as the original variable type, its array dimensions, the precise time it has been shared, the host and/or program that provided the data, and statistics on how may times the data has been updated or accessed. For example, the original data type of field {\ttfamily bar} in table {\ttfamily foo}, is stored in the Redis hash table $\langle${\ttfamily types}$\rangle$ under the field named {\ttfamily foo:bar}. See Table~\ref{tab:metadata} for details. The essential metadata is added / updated every time data is written into SMA-X via the C or Python libraries, command-line tools, or directly via the Lua helper scripts. And, it is retrieved each time data is read from SMA-X via the libraries, command-line tools, or Lua helpers.  Beyond the essential metadata, additional optional metadata may be provided as needed (see Table~\ref{tab:optional-metadata}). The optional metadata must be explicitly provided by the producer program, if desired, usually through an explicit library call. The producer may provide these only once, e.g. at first write after program start, or update them each time the data is updated -- whichever is appropriate for the particular type of extra metadata. Similarly, consumers of optional metadata must retrieve these separately, once or every time time, as appropriate. In the future we may extend the SMA-X standard to include further types of optional metadata also.

\begin{table}[htb]
\caption{Essential metadata, stored for every variable in SMA-X.}
\centering
\begin{tabular}{|c|l|}
\hline
 {\bf Meta Table} & {\bf Description} \\
\hline
$\langle${\ttfamily types}$\rangle$      & storage data type of each variable, see Table~\ref{tab:data-types}. \\
\hline
$\langle${\ttfamily dims}$\rangle$       & array dimensions for each variable. \\
\hline
$\langle${\ttfamily timestamps}$\rangle$ & precision UNIX timestamps when the variable was last updated in SMA-X. \\
\hline
$\langle${\ttfamily origins}$\rangle$    & host and/or program that provided the variable \\
\hline
$\langle${\ttfamily reads}$\rangle$      & number of times the variable as read by clients \\
\hline
$\langle${\ttfamily writes}$\rangle$     & serial number, i.e.\ number of times the variable was updated in SMA-X \\
\hline
\end{tabular}
\label{tab:metadata}
\end{table}

\begin{table}[htb]
\caption{Optional metadata, stored and accessed as needed.}
\centering
\begin{tabular}{|c|l|}
\hline
 {\bf Meta Table} & {\bf Description} \\
\hline
$\langle${\ttfamily descriptions}$\rangle$ & a concise description of what data the variable stores \\
\hline
$\langle${\ttfamily units}$\rangle$        & physical unit of the data (e.g.\ if not SI). \\
\hline
$\langle${\ttfamily coords}$\rangle$       & coordinate system description for array data in one or more dimensions \\
\hline
\end{tabular}
\label{tab:optional-metadata}
\end{table}

\subsubsection{Hierarchical data} 
\begin{figure}[!hbt]
\includegraphics[width=\textwidth]{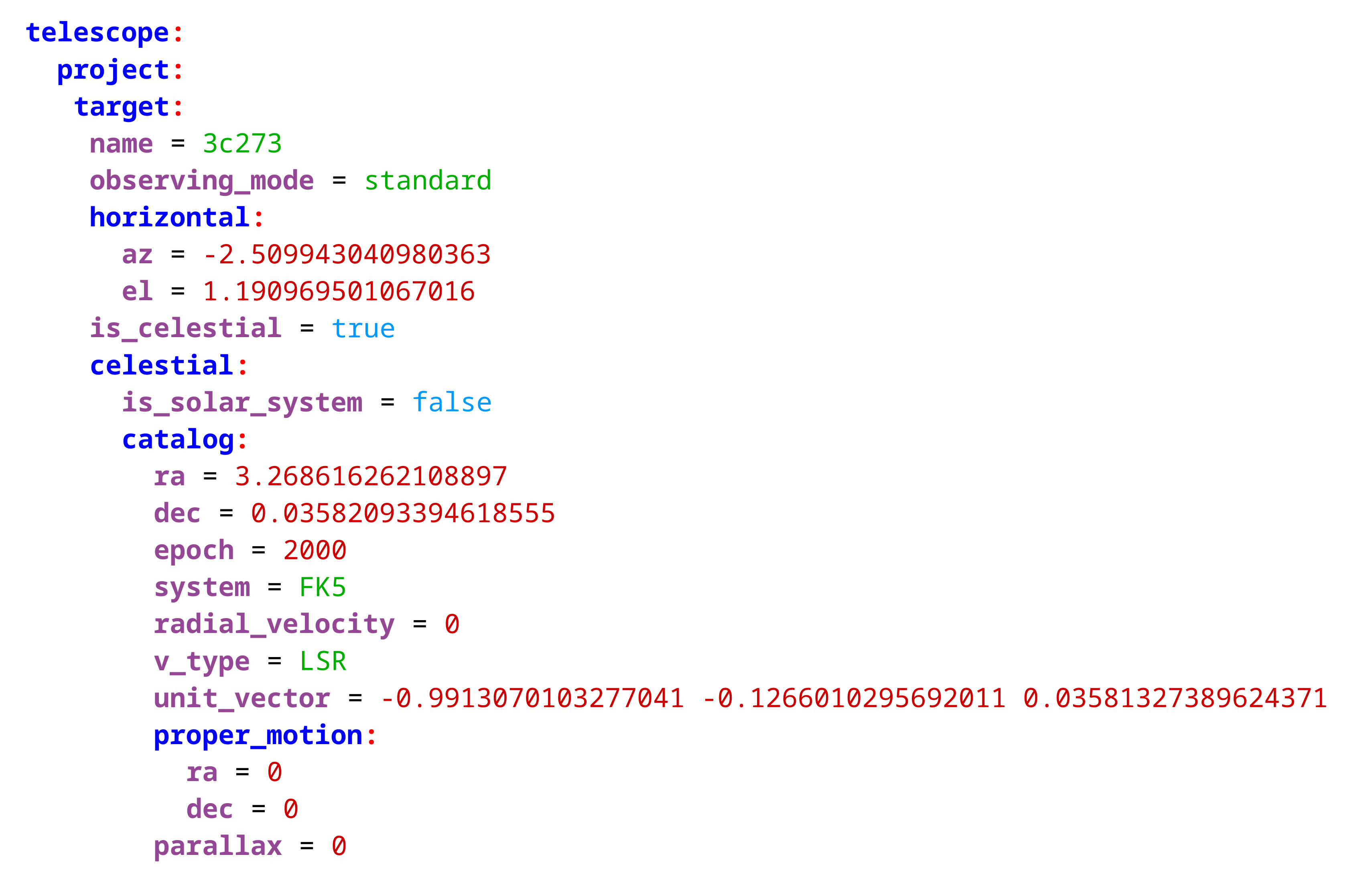}
\caption{\label{fig:example} Example structured data layout in SMA-X. This excerpt stores data about the currently observed source in the project. For example the source name is stored in the Redis hash table \texttt{telescope:project:target} under the field \texttt{name}, and has the string value of \texttt{3c273}. The values are shown as they appear in the database in the serialized form as strings.}
\end{figure}

SMA-X is designed to store data organized hierarchically, much a like a file-system would (see Figure~\ref{fig:example}). It allows data to be grouped in a logical way by systems, subsystems, components etc, so they may be easily located and related bits of information can be accessed together easily. Just like a file-system layout, the hierarchy is decided by the program that produces the data and writes it into SMA-X. The consumers then simply get to view that data in the form inteded by the producer (with some added flexibility on casting leaf nodes to different types, and/or padding or truncating to different elements counts). The SMA-X storage convention allows for efficient atomic read/write access of structured data both as a whole, or any of its branches or leaf-nodes individually. The level of access provided by SMA-X is generally similar to the capabilities of the built-in Redis Javascript Object Notation (JSON)\cite{JSON, JSON-ISO} support, which was introduced in Redis 7 after the inception of SMA-X. However, SMA-X also offers features that extend beyond the native JSON support with its support for metadata, such timestamps etc.

\subsubsection{Atomicity} 
Bundled data, such as the included metadata and/or structures, are both written to and retrieved from the SMA-X database in a single atomic operation. Thus, the user need not worry about concurrency issues, such as one client getting incompletely updated data while another is in the middle of setting new values for these. This will never happen with SMA-X as long as the user reads and writes bundled data with singular calls. (In the current version of SMA-X, atomicity only applies for each layer at a time of deep structures when writing, but it is a limitation we hope to overcome soon via a new Lua script and/or function as well as corresponding changes to the C and Python libraries.)

\subsubsection{Update notifications} 
Every time a variable or entire structure in SMA-X is updated, a set of notifications on specific PUB/SUB channels bearing the variable's name (as well as those of its parent hierarchies) are sent, so that subscribed clients are notified immediately of state changes on any/all variables of interest, and can act on these as appropriate. This is similar to the \texttt{INVALIDATE} push messages introduced in Redis 7 with it \texttt{CLIENT TRACKING} feature, but SMA-X will send these update notifications not only for the leaf nodes that changed, but for its parent hierarchies, also -- which is critical for the robust detection state changes for hierarchical data. For example, if \texttt{telescope:project:target:name} is changed, SMA-X will also notify that the embedding structures \texttt{telescope:project:target}, \texttt{telescope:project}, and \texttt{telescope} have changed as a result also. (And, if multiple fields are updated at once inside a structure, SMA-X will send just the necessary single notification for the parents at the end of the update.)

\subsubsection{Remote settings} 
Extending on the set of features described above, we also specify a convention for 'commanding' program settings remotely through SMA-X. Remote settings are analogous to Remote Procedure Calls (RPC), except that they do not provide a 'computed' return value for the specific caller. (You may think of them as RPC with {\ttfamily void} return type.) The SMA-X database simply stores identical sets of {\ttfamily commanded} and {\ttfamily actual} values that represent a program's settings. The program may monitor the {\ttfamily commanded} set for changes, and then report applied values back in the {\ttfamily actual} set, which in turn a client program may monitor for confirmation that changes were applied (and how exactly).

\subsubsection{Remote messaging} 
Beyond the PUB/SUB channels that carry notifications for all SMA-X variables, we also standardize the naming for channels that may carry messages from individual programs, and which can be monitored by clients, e.g.\ to selectively monitor progress or errors / warnings remotely.

\subsection{SMA-X Client Tools \& Libraries}

We provide C/C++ and Python 3 libraries, as well as a simple set of command-line tools to help access and maintain your custom SMA-X database. The command-line tool {\ttfamily smaxValue} can be used to query SMA-X variables, including the essential metadata, or to list the contents of specific tables, while {\ttfamily smaxWrite} can be used to add/update SMA-X data from the command-line. Both command-line tools can be used with scripting languages, such as {\ttfamily bash}, {\ttfamily perl} or similar.

The C/C++ and Python libraries come with generally similar set of features. Python, of course, allows for a cleaner, type-agnostic interface, and for exception handling. In contrast, the C/C++ library offers families of related functions, with separate calls providing the same functionality for each supported data type, and it relies on return values and {\ttfamily errno} to indicate error states. Both libraries offer detailed API documentation to guide users.

Below we summarize some of the main features of the C/C++ library, noting here only that some of these may be implemented somewhat differently for the Python client library. Refer to the documentation included with the libraries themselves on their particular implementations.

The SMA-X C/C++ client keeps up to three separate TCP/IP channels open to the SMA-X Redis server: (1) for interactive transactions; (2) for pipelined (batch) requests; and (3) for PUB/SUB notifications and management thereof.

\subsubsection{Automatic type conversions} 
The C/C++ library implements automatic type conversions for leaf-node access, such that the client software can access the shared data as a different type from how it was produced. Because data is always stored as string(s), the parse type of data may be anything really, regardless of the original storage type, as long as the stored string value(s) can be parsed as the desired type. This includes both widening and narrowing conversions, such as from {\ttfamily int16} to {\ttfamily int8}, {\ttfamily int64}, or {\ttfamily float}, and vice-versa -- as well as conversions between numerical, logical, and string types also. The Python library is intrinsically not strongly typed and allows for most of the same conversions with little or no effort. This feature allows tolerating changes to data type that is generally transparent to dependent clients. Say you used a 32-bit floating point value to store a measurement, and have written clients to process data as 32-bit \texttt{float}. One day you decide that to you want to store data with higher precision, so you start writing 64-bit double-precision values instead. Your existing clients will continue to read the data as 32-bit \texttt{float}, without breaking their functionality.

\subsubsection{Automatic array padding / truncation} 
The C/C++ library implements automatic padding or truncation to serve the user application the data it wants / needs. Consider an array of values that contain a set of readings from $N$ instruments or channels. You have written a client application processing these $N$ values. Then, new hardware is deployed, adding another $M$ units / channels to the system, so the SMA-X database now has $N+M$ elements in the array for the same variable. The C library will continue to deliver just the $N$ values for existing clients that they expect. It does not break anything if the client does not process the additional values. Of course, you may want to update the client application sooner or later to process $N+M$ values, but in the meantime, it will keep chugging on the 'old' way. Similarly, if some units are decommissioned, and fewer than $N$ values are reported in the SMA-X array data, the C library will just pad the remaining elements with zeroes so your application can continue to process $N$ values the same as always.

\subsubsection{Adapting with metadata} 
While the above features allow existing clients to continue 'normal' operation even if the data in SMA-X may have changed it type or element count, it may not be the best or the desired way for the client to adapt to changes. If you anticipate that the type or size of some leaf nodes may change as the telescope ecosystem evolves, you can make your client applications more future proof by using the metadata to check the actual data type and size, as was written, and then use the data accordingly. This way, clients can detect changes in data type or element count on the fly, and properly adapt their behavior accordingly. The C/C++ library provides functions to return current values in dynamically sized arrays also, much like the Python API readily does. We also note, that the addition of new nodes or structures into SMA-X has absolutely no effect on existing clients and functionality. Existing SMA-X clients can continue operating uninterrupted as the database is expanded to contain new nodes.

\subsubsection{Pipelining (batch mode)}
Data that is pushed to SMA-X is sent out via the pipelined connection (provided it's available). As such, data sharing from your application is always non-blocking and incurs no unexpected overheads, since it does not wait for confirmation back from the Redis server. This fire-and-forget approach ensures that sharing of information is safe even when used from within timing critical processes and threads. Data can also be retrieved asynchronously in batches through the pipeline connection. Queries can be submitted to a queue, and the responses that are received for these are processed asynchronously in a separate background thread. The user or application can perform other time critical tasks while the responses are gathered, and then either wait until the requested batch of data is available (via appropriate synchronization points inserted into the queue), or else request an asynchronous callback. Because pipelined transfers do not require a chain of round-trips to the SMA-X server over the network, they can provide orders of magnitude higher throughput for bulk transfers than sequentialized individual interactive transactions. 

\subsubsection{Lazy access} 
The SMA-X library provides lazy access for data retrieval. Lazy retrieval is most useful when one needs information more frequently than it is produced, without flooding the network with frequent polling requests. It essentially amounts to maintaining locally cached copies of the requested SMA-X variables. The caches are either invalidated if SMA-X notifies of an update, or else seamlessly updated in the background. (The caching mode can be selected for each lazy variable individually if desired.) When your application needs the current value of a variable, it is either retrieved instantly from the local cache (with zero network traffic), provided that the cache is valid, or else its current value is fetched from the SMA-X database.

\subsubsection{Semaphores \& callbacks} 
We provide means to wait (block the current thread) until some variable or some set of variables change. Monitored variables to wait upon can be designated individually or via glob patterns, and a timeout value can also be specified to return with an error (or exception) in case no changes were detected in the allotted amount of time. Alternatively, one may specify a callback function to be invoked asynchronously when a variable with matching name or pattern is updated.

\subsubsection{Resiliency} 
We designed our C/C++ and Python libraries to be generally resilient to intermittent outages of the SMA-X server or the network infrastructure. If locally produced data cannot be pushed to the SMA-X server, the libraries will maintain a local store of the pending updates, keeping only the most recent values for any shared variable, and will try reconnecting to the SMA-X server in the background at regular intervals. If and when SMA-X is successfully reconnected, the pending local updates are pushed before any new data is sent to or received from the SMA-X server. Note, that since the local store will cache only the most current version of each 'sent' variable, its storage needs will not grow with repeated submissions for the same variable while the server connection is down, and the total size of the local store will never exceed the storage size of the data produced locally, notwithstanding some additional overhead.

\subsection{Performance Notes}

We benchmarked SMA-X on an AMD Ryzen 5 PRO 6650U (released in 2022) laptop running Fedora Workstation 41 (kernel 6.12.4), which acted as both client and server for the tests, using the loopback interface, and Valkey 8.0.1 for the SMA-X database backend, and the \texttt{smax-clib} C library for the client (the benchmarking program is included in the \texttt{smax-clib} GitHub repository). It's not necessarily the most ideal, or cutting edge benchmarking setup, but it provides a ballpark idea for the level of peak performance that you might expect on fast Local Area Network (LAN) also. 

In this setup, we could consistently write ~75,000 variables (including metadata) per second, and read $\sim$205,000 variables (also with metadata) per second for assorted types (a collection of scalars of various types, small arrays, and strings) in pipelined mode, or read around 33,000 variables per second in interactive (round-trips) mode.

The interactive mode performance is likely limited by the roundtrip time, which was measured around 30--70\,$\mu$s with the \texttt{ping} command. Thus, the 33,000 round-trip transactions per second are broadly consistent with the expected limits of the loopback interface itself. The pipelined mode throughput is likely limited by the Valkey server performance. Here we note that the use of Lua scripts seems to have only a moderate effect on the overall performance, which for retrievals remained within a factor of $\sim$1.5 of the performance measured for raw Redis \texttt{HGET} command in the same test setup and for the same data, without the metadata.

Similar tests performed on the SMA telescope 1\,GB/s network, between two computers (a physical machine and a VM), yielded lower metrics at around 40,000 writes per second, or around 48,000 pipelined reads per second, and interactive performance at 2,600--3,000 transactions per seconds (in excellent agreement with the \texttt{ping} roundtrip timing measurements at around 400\,$\mu$s). What this shows, is that in a real-world applications, the true performance of SMA-X is likely limited by the latencies and the throughput of the telescope network, and not so much by the software layer itself.

\subsection{Creating a Historical Record}

By nature SMA-X stores only a snapshot of the current state of all shared variables. It has no sense of history or prior state information, by design. (Redis offers such features, but SMA-X does not use these.) It is however often useful to keep a historical record of the observatory's runtime state. It can help diagnose problems after they occur; or validate data at a later time; and engineers may use it to troubleshoot hardware and/or software components by having access to a long-term behavioral record. At the SMA we have been keeping such a historical record of shared data (even before SMA-X) at 1-minute resolution, going back 12 years. We find it an invaluable resource, and the possibility of being able to check specific conditions and sensor values at the telescope on a given day and time some years back is a priceless feature.

For this reason, we have also developed a connector daemon which can regularly sample and/or snapshot all, or a selection of, SMA-X variables at regular intervals and store them in a PostreSQL database\cite{Postgres}, with or without a TimescaleDB extension for more efficient retrieval of the time-ordered data. At the Submillimeter Array, we store a snapshot of nearly all SMA-X variables once every hour, and store updates to changing variables once per minute as necessary. The connector application stores metadata history also, when changes to metadata are detected. The connector application is highly configurable. You can set:

\begin{itemize}
\item{the database location, name, and user credentials.}
\item{whether or not TimescaleDB extension should be used.}
\item{the frequency of regular updates for changing variables (e.g.\ 1 minute).}
\item{the frequency of full snapshots of the SMA-X database (e.g.\ 1 hour).}
\item{variables and glob patterns specifying which variables to include in or exclude from the archival.}
\item{a maximum size for variables that will be archived automatically.}
\item{a maximum age of variables to be included automatically (so that orphaned data is not polluting the historical record forever).}
\item{a sampling for array variables where only every $n^{\rm th}$ array element is stored.}
\item{to force archival of select variables or patterns even if they would otherwise be excluded by one of the other settings above.}
\end{itemize}

The connector daemon can be integrated with and managed via {\ttfamily systemd}, e.g.\ to start automatically after boot. It will monitor the SMA-X database continuously, and insert new tables into the PostgreSQL database as necessary for any new variable(s) appearing in SMA-X, provided these aren't excluded from archiving by the configuration.

A particularly welcome aspect of having a historical record stored in a PostgreSQL database, is that it is very easy to create time-series visualizations of these, e.g.\ with Grafana. As such, one can easily produce web pages for visual monitoring of shared variables of particular importance.

\subsection{Current Status and Future Plans}

SMA-X has been in use at the Submillimeter Array since 2021, where it has proven to be highly stable and reliable. We use both the C and Python libraries routinely for our everyday operation. SMA-X is also used with the new control system of the Massachussetts Institute of Technology (MIT) Haystack 37-m telescope\cite{Haystack37}. The SMA-X source code, libraries, and tools are open-source and freely available to anyone via the set of GitHub repositories listed in Table~\ref{tab:repos}.

\begin{table}[htb]
\caption{SMA-X GitHub repositories.}
\centering
\begin{tabular}{|l|l|}
\hline
 {\bf GitHub repository} & {\bf Description} \\
\hline
{\ttfamily Smithsonian/smax-server}           & server configuration and {\ttfamily systemd} integration. \\
\hline
{\ttfamily Smithsonian/smax-clib}             & C/C++ client library and command-line tools. \\
\hline
{\ttfamily Smithsonian/smax-python}           & Python3 client library. \\
\hline
{\ttfamily Smithsonian/smax-postgres}         & PostgreSQL connector application \\
\hline
\end{tabular}
\label{tab:repos}
\end{table}

There is still room for SMA-X to grow in capabilities. We readily foresee a number of avenues for adding new features or improving existing ones. At present we are considering the following areas, in which SMA-X can extend and expand functionality:

\begin{itemize}
\item{Fully atomic deep structure writes.}
\item{Initializing optional metadata and static configuration data via a separate persistent configuration database such as a MongoDB or an SQL database. This way SMA-X may provide the same access protocol for static data, which is not directly produced by the real-time system.}
\item{More optional metadata, such as normal operating ranges, and critical levels for measured values, which may be used e.g.\ by an automated warning system.}
\item{Direct support for complex-valued data types (e.g. {\ttfamily complex[32,64]}).}
\item{Standardized support for non-persistent, streaming-only data via PUB/SUB. It may be useful to provide bundled, self-contained, data packets (e.g. real-time telescope pointing information) published at a fast cadence ($>$10\,Hz) to support low-latency real-time client applications for these.} 
\item{Use Redis list storage types to implement data queues (FIFOs) on SMA-X, such as for managing an observing queue, or a set of tasks that have to be executed sequentially.}
\item{Asynchronous Remote Procedure Call (RPC) through SMA-X. SMA-X could allow for more proper zeroconf remote calls using the Redis PUB/SUB infrastructure, with unique program IDs (e.g.\ {\ttfamily host:program[:task]}) in lieu of physical addresses. We have a draft protocol for RPC over SMA-X, but it has yet to be finalized, and implemented for the client libraries.}
\item{Client libraries for other languages, e.g.\ Java, Rust, Go. We welcome external contributions for implementing additional client support for the programming language(s) of your choice.}
\item{We plan to package the SMA-X server, libraries, and tools for Linux distributions (e.g. Debian and Fedora RPM packages), as well as for Homebrew (MacOS X) to make it more easily available to users and applications.}
\item{Enhanced security. SMA-X presently relies on restricting access to internal networks to prevent unauthorized access. Redis provides additional security features, such as database user authentication, SSH-tunnel-only access, and Transport Layer Security (TLS) support -- which can enhance and further protect the database from outside attacks.}
\item{Let us know what other added feature would make SMA-X work better for you.}
\end{itemize}

\section{Conclusion}

SMA-X offers an appealing, fast, and versatile solution for distributed monitoring and control in and around telescopes, or for other distributed systems. It relies on a central Redis server, as a public message board within the telescope ecosystem, for sharing and accessing structured data to/from various programs and nodes. Data access is atomic both for arbitrary data branches, sub-branches, or leaf elements, and includes metadata that describe the stored values. SMA-X unites persistent storage of the current state of shared information with a PUB/SUB notification system for updates -- a powerful combination which allows clients to efficiently access the information both on demand and/or when values are updated. The dual mode information sharing minimizes network traffic, and eliminates latencies beyond those of the underlying networking layer. The centralized Redis server also minimizes the number of network connections necessary to $\mathcal{O}(N)$ instead of the typical $\mathcal{O}(N^2)$ links required in distributed P2P systems. And, because all information is available to all programs/nodes within the ecosystem, SMA-X supports distributed decision making in a way P2P typically does not. As such SMA-X provides a superior solution to existing P2P models for data sharing in distributed systems in general.

We provide a set of C/C++ and Python libraries that client applications can use to bring the most out of SMA-X. These libraries offer additional features, such as asynchronous pipelining (fast bulk data access), lazy access (with or without background caching), semaphores and callbacks that can be used to act immediately on specific state changes, and resiliency for intermittent network or server outages. Simple command-line tools provide basic access to SMA-X that can be used with any scripting language (e.g.\ {\ttfamily bash} or {\ttfamily perl}) also. Furthermore, we provide a configurable application that can be used to create a long-term historical record of all shared state variables in a PostgresSQL database. These may be visualized, e.g. with Grafana, to show time evolution of select monitoring states, or can be used for diagnostics in general. (At the SMA, we have such a record going back 12 years with a 1 minute cadence, predating even SMA-X -- and we find it enormously useful for diagnostics, and for data quality related checks even years after the fact).

It is our plan to maintain and develop SMA-X for the foreseeable future. As such, we expect to expand and enhance the capabilities it provides, and evolve the client libraries we maintain to match. All of our existing SMA-X related code has been open-sourced, and we hope to offer packaged versions in the near future also. We will welcome your questions, comments, suggestions, or any other feedback that you may provide to make SMA-X better.

\subsection* {Disclosures}

The authors declare that there are no financial interests, commercial affiliations, or other potential conflicts of interest that could have influenced the objectivity of this research or the writing of this paper.

\subsection* {Code, Data, and Materials Availability} 

All SMA-X related source code, configuration files, and documentation is publicly accessible via the set of GitHub repositories listed in Table~\ref{tab:repos}.

\subsection*{Acknowledgements}

The Submillimeter Array is a joint project between the Smithsonian Astrophysical Observatory and the Academia Sinica Institute of Astronomy and Astrophysics and is funded by the Smithsonian Institution and the Academia Sinica.

A previous version of this paper was published in the Proceedings of the SPIE Conference 13101\cite{spie-paper}.

\bibliography{smax} % bibliography data in report.bib
\bibliographystyle{spiejour} % makes bibtex use spiebib.bst

\pagebreak

\vspace{2ex}\noindent\textbf{Attila Kov\'{a}cs} is a computer engineer at the Center for Astrophysics $|$ Harvard \& Smithsonian. He received his A.B. degree in physics, astronomy and astrophysics from Harvard College in 1997, and his PhD degree in physics from the California Institute of Technology in 2006. His current research interests include submillimeter detector technologies and instrumentation, star-forming galaxies, signal processing, and data reduction and imaging algorithms.

\vspace{2ex}\noindent\textbf{Paul K. Grimes} received the B.A. (Hons) and M.Sci. degrees in experimental and theoretical physics from King's College, University of Cambridge, Cambridge, U.K., in 2001, and the Ph.D. degree in astrophysics from the Cavendish Laboratory, University of Cambridge, in 2006. From 2005 to 2011, he was a Postdoctoral Researcher with Oxford Astrophysics, University of Oxford, Oxford, U.K., developing astronomical instrumentation for radio, millimeter and submillimeter-wave astronomy, and cosmology, also carrying out research on superconducting bolometers and heterodyne mixers. In 2011, he joined the Smithsonian Astrophysical Observatory, Cambridge, MA, USA, as a Physicist. His main research interest includes the development of superconducting mixers and other instrumentation for submillimeter-wave astronomical telescopes.

\vspace{2ex}\noindent\textbf{Christopher Moriarty} is a cloud engineer and part-time contractor to the Minor Planet Center since August 2024. Prior to that he was a Senior Software engineer for the SMA (2018--2021) and then Technical Manager (2021--2024) at the Center for Astrophysics $|$ Harvard \& Smithsonian.

\vspace{2ex}\noindent\textbf{Robert Wilson} is a senior scientist at the Center for Astrophysics $|$ Harvard \& Smithsonian (1994-present). He co-discovered cosmic microwave background radiation and interstellar CO with Arno Penzias. His other institutional affiliations include Bell Laboratories and California Institute of Technology.

%\listoffigures
%\listoftables

%\end{spacing}
\end{document}